\documentclass[aip,preprint]{revtex4-1}
\usepackage{setspace}
\usepackage{hyperref}
\usepackage{natbib}
\usepackage{graphicx}
\usepackage{siunitx}
\usepackage{threeparttable,diagbox,multirow}
\usepackage{xcolor}
\usepackage{amsmath}
\usepackage{braket}
\usepackage{mhchem}

\draft 
\begin{document}

\title{Improving the efficiency of $G_0W_0$ calculations with approximate spectral decompositions of dielectric matrices}

\author{Han Yang}
\affiliation{Department of Chemistry, University of Chicago, Chicago, Illinois 60637, United States}
\affiliation{Pritzker School of Molecular Engineering, University of Chicago, Chicago, Illinois 60637, United States}

\author{Marco Govoni}
\affiliation{Pritzker School of Molecular Engineering, University of Chicago, Chicago, Illinois 60637, United States}
\affiliation{Materials Science Division and Center for Molecular Engineering, Argonne National Laboratory, Lemont, Illinois 60439, United States}

\author{Giulia Galli}
\email[]{gagalli@uchicago.edu}
\affiliation{Department of Chemistry, University of Chicago, Chicago, Illinois 60637, United States}
\affiliation{Pritzker School of Molecular Engineering, University of Chicago, Chicago, Illinois 60637, United States}
\affiliation{Materials Science Division and Center for Molecular Engineering, Argonne National Laboratory, Lemont, Illinois 60439, United States}

\date{\today}

\begin{abstract}
Recently it was shown that the calculation of quasiparticle energies using the $G_0W_0$ approximation can be performed without computing explicitly any virtual electronic states, by expanding the Green function and screened Coulomb interaction in terms of the eigenstates of the static dielectric matrix. Avoiding the evaluation of virtual electronic states leads to improved efficiency and  ease of convergence of $G_0W_0$ calculations. Here we propose a further improvement of the efficiency of these calculations, based on an approximation of density-density response functions of molecules and solids. The approximation relies on the calculation of a subset of eigenvectors of the dielectric matrix using the kinetic operator instead of the full Hamiltonian, and it does not lead to any substantial loss of accuracy for the quasiparticle energies.  The computational savings introduced by this approximation depend on the system, and they become more substantial as the number of electrons increases.
\end{abstract}


\maketitle 

\section{\label{sec:introduction}Introduction}
	Devising accurate and efficient methods to predict the electronic properties of molecules and condensed systems is an active field of research. Density functional theory (DFT) has been widely used for electronic structure calculations.\cite{DFT1964,DFT1965,DFT2004} However, the exact form of the exchange-correlation functional is unknown and therefore DFT results depend on the choice of approximate functionals. Improvement over DFT results may be obtained by using many-body perturbation theory (MBPT).\cite{ReiningMBPT,martin2016InteractingElectrons} A practical formulation of MBPT for many electron systems was proposed by Hedin,\cite{hedin1965} where the self-energy $\Sigma$ is written in terms of the Green's function $G$ and the screened Coulomb interaction $W$.

	The  $GW$ approximation\cite{hedin1965,martin2016InteractingElectrons} has been successful in the description of the electronic properties of several classes of materials and molecules;\cite{OriginalGW100,GW100-maggio2017,Marco2018,peter2016JCTC,brawand2016generalization,golze2019gw-compendium} however the computational cost of $GW$ calculations remains rather demanding and many complex systems cannot yet be studied using MBPT. Hence, algorithmic improvements are required to apply MBPT to realistic systems. One of the most demanding steps of the original implementation of $GW$ calculations\cite{Louie1985PRL,Louie1985PRB,Louie1986,Strinati1980GWImplementation,Strinati1982GWImplementation} involves an explicit summation over a large number of unoccupied single particle electronic orbitals, which enter the evaluation of the dielectric matrix $\epsilon$\cite{adler1962quantum,wiser1963dielectric} defining the screened Coulomb interaction $W$ ($W=\epsilon^{-1}v_c$, where $v_c$ is the Coulomb interaction). The summation usually converges slowly as a function of the number of virtual orbitals ($N_c$).

	In recent years, several approaches have been proposed to improve the efficiency of $GW$ calculations. For example, in Ref. \citenum{Louie2011PRL} it was suggested to replace unoccupied orbitals with approximate physical orbitals (SAPOs); the author of Ref. \citenum{ExtrapolarApprox} simply truncated the sum over empty states entering the calculation  of the irreducible density-density response function, and  assigned the same,  average energy to all the empty states  higher than a preset value; in a similar fashion, in Ref. \citenum{PeihongZhang}  an integration over the density of empty states  higher than a preset value was used. Other approaches adopted sophisticated algorithms to invert the dielectric matrix, e.g. in Ref. \citenum{Multi-Pole-Approx}, they employed a Lanczos algorithm.
	Recently, an implementation of $G_0W_0$ calculations avoiding altogether explicit summations over unoccupied orbitals, as well as the necessity to invert dielectric matrices, has been proposed,\cite{wilson2008,wilson2009,Nguyen2012-GW,Anh2013-GW,Marco2015} based on the spectral decomposition of density-density response functions in terms of eigenvectors (also known as projective dielectric eigenpotentials, PDEPs). In spite of the efficiency improvement introduced by such formulation, $G_0W_0$ calculations for large systems remain computationally demanding.

	In this paper, we propose an approximation to the projective dielectric technique,\cite{wilson2008} which in many cases leads to computational savings of $G_0W_0$ calculations of 10-50\%, without compromising accuracy. The rest of the paper is organized as follows: we describe the proposed methodology in \autoref{sec:method} and then we present results for several systems in \autoref{sec:results}, followed by our conclusions.

\section{\label{sec:method}Methodology}
We compute the density-density response function of solids and molecules within the framework of the random phase approximation (RPA), using  projective dielectric eigenpotentials (PDEP)\cite{wilson2008,wilson2009,Marco2015}. The accuracy of this approach has been extensively tested for molecules and solids.\cite{wilson2009} The technique relies on the solution of the Sternheimer's equation~\cite{Sternheimer1954,DFT2004,Galli1993}
\begin{equation}
    (\hat{H}-\varepsilon_v\hat{I})\lvert\Delta\psi_v\rangle = -\hat{P}_c\Delta \hat{V}\lvert\psi_v\rangle
    \label{equ:Sternheimer}
\end{equation}
to obtain the linear variation of the $v$-th occupied electronic orbital, $\lvert\Delta\psi_v\rangle$, induced by the external perturbation $\Delta \hat{V}$. In \autoref{equ:Sternheimer}, $\hat{I}$ is the identity operator, $\hat{P}_c$ is the projector onto the unoccupied states, $\varepsilon_v$ and $\psi_v$ are the $v$-th eigenvalue and eigenvector of the  unperturbed Kohn-Sham Hamiltonian $\hat{H}=\hat{K}+\hat{V}_\mathrm{SCF}$, respectively, where $\hat{K}=-\frac{\nabla^2}{2}$ is the kinetic energy, $\hat{V}_\mathrm{SCF}$ is the self-consistent potential.
For each perturbation, the first order response of the density $\Delta n$ can be obtained as \cite{baroni2001phonons}
\begin{equation}
    \Delta n = 2 \sum_v \psi_v\Delta\psi_v +c.c. \,
    \label{eq:deltadensity}
\end{equation}
\autoref{equ:Sternheimer} and \ref{eq:deltadensity} can be used to iteratively diagonalize the static symmetrized \textit{irreducible} density-density response, $\tilde{\chi}_0$:\cite{wilson2008,wilson2009,Marco2015}
\begin{equation}
    \tilde{\chi}_0 = \sum_{i=1}^{N_\mathrm{PDEP}}\left|\xi_i\right\rangle\lambda_i\left\langle\xi_i\right|,\label{equ:chi0}
\end{equation}
where $\lambda_i$ and $\xi_i$ are eigenvalues and eigenvectors of $\tilde{\chi}_0$ and $N_\mathrm{PDEP}$ is the number of eigenvectors of $\tilde{\chi}_0$, respectively. The eigenvectors $\xi_i$ are called PDEPs: projective dielectric eigenpotentials throughout the manuscript. The symmetrized irreducible density-density response function is defined as $\tilde{\chi}_0 = v_c^{1/2}\chi_0v_c^{1/2}$, where $v_c$ is the Coulomb potential.\cite{Marco2015} Within the RPA, the symmetrized \textit{reducible} density-density response can be expressed as  $\tilde{\chi}=\left(1-\tilde{\chi}_0\right)^{-1}\tilde{\chi}_0$, therefore the $\xi_i$ are also eigenvectors of $\tilde{\chi}$. The projective dielectric technique has also been recently applied beyond the RPA using a finite field method.\cite{HeMa2018JCTC}

When solving the Sternheimer's equation, it is not necessary to compute explicitly the electronic empty states, because one can write $\hat{P}_c = \hat{I}-\hat{P}_v$, since the eigenvectors of $\hat{H}$ form a complete basis set ($\hat{P}_v$ is the projector onto the occupied states). The use of \autoref{equ:chi0} significantly reduces the cost of $G_0W_0$ calculation from $N_\mathrm{pw}^2N_vN_c$ to $N_\mathrm{PDEP} N_\mathrm{pw}N_v^2$ where $N_v$, $N_c$, $N_\mathrm{PDEP}$, $N_\mathrm{pw}$ are numbers of occupied orbitals (valence bands in solids), virtual orbitals (conduction bands in solids), PDEPs and plane waves, respectively, and importantly $N_\mathrm{PDEP}\ll N_\mathrm{pw}$.

\begin{figure}
    \centering
    \includegraphics[width=0.6\linewidth]{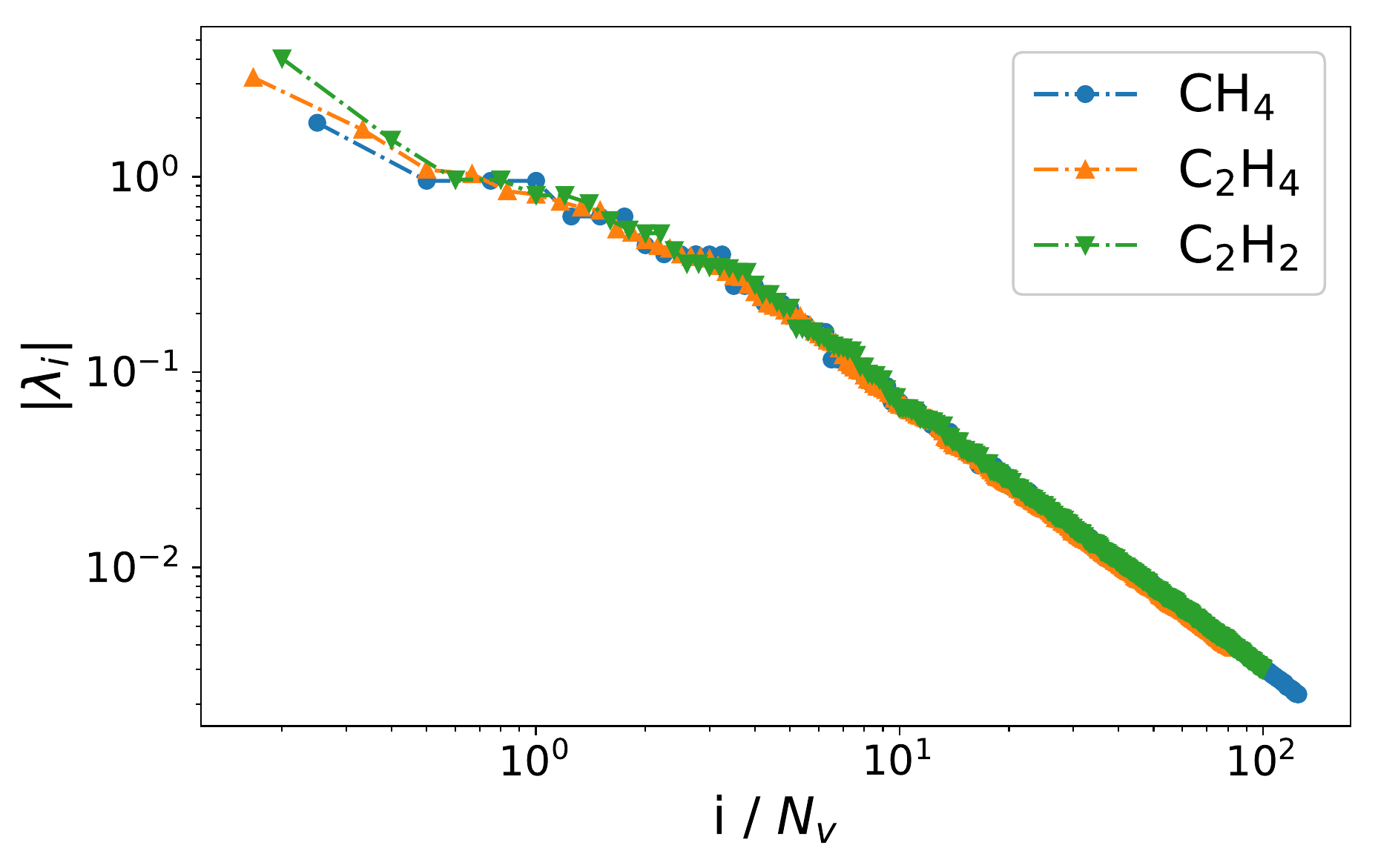}
    \caption{First 500 eigenvalues $\lambda_i$ of the symmetrized irreducible density-density response function $\tilde{\chi}_0$ (see text), for three small molecules: \ce{CH4} (blue dots), \ce{C2H4} (orange up triangles), and \ce{C2H2} (green down triangles). $N_v$ is the number of occupied orbitals.}
    \label{fig:wstat-eigens}
\end{figure}

The application of the algorithm described above   to large systems is hindered by the cost of  solving \autoref{equ:Sternheimer}. However, we note that the eigenvalues of $\tilde{\chi}_0$ rapidly converge to zero ,\cite{Deyu2008PRL,wilson2008,wilson2009,Marco2018} (an example is shown in \autoref{fig:wstat-eigens}). In addition, as shown in Ref. \citenum{wilson2009}, the eigenvalue spectrum of the dielectric function for eigenvectors higher than the first few, is similar to that of the Lindhard function.\cite{LindhardFunction1954} Hence we propose to compute the PDEPs of $\tilde{\chi}_0$ corresponding to the lowest eigenvalues using \autoref{equ:Sternheimer} and \ref{eq:deltadensity} and to compute the remaining ones with a less costly approach. Inspired by the work of \citet{Rocca2014}, we  approximate the eigenpotentials corresponding to higher eigenvalues with kinetic eigenpotentials, which are obtained approximating the full Hamiltonian entering \autoref{equ:Sternheimer} with the kinetic operator($\hat{K}$):\cite{RPA-total-energy-PRL2009,Rocca2014}
\begin{equation}
    (\hat{K}-\varepsilon_v\hat{I})\lvert\Delta\psi_v\rangle = -\hat{P}_c\Delta \hat{V}\lvert\psi_v\rangle\label{equ:KineticSternheimer}.
\end{equation}
We note that the application of the kinetic energy ($\hat{K}$) amounts simply to computing a sum over the plane-wave expansion coefficients multiplied by the square of the $\mathbf{G}$-vectors (we are using a plane-wave basis set). Numerous additional operations involving the solution of a Poisson equation and integrals in real space are necessary to apply the full Hamiltonian in \autoref{equ:Sternheimer}, which includes the Hartree and exchange correlation potentials. Furthermore, the evaluation of the non-local pseudopotentials (not needed when solving \autoref{equ:KineticSternheimer}) is an expensive operation in principle of $O(N^3)$, where $N$ is the number of electrons.\cite{DFT2004,Galli1993}

\begin{figure}
    \centering
    \includegraphics[width=0.8\linewidth]{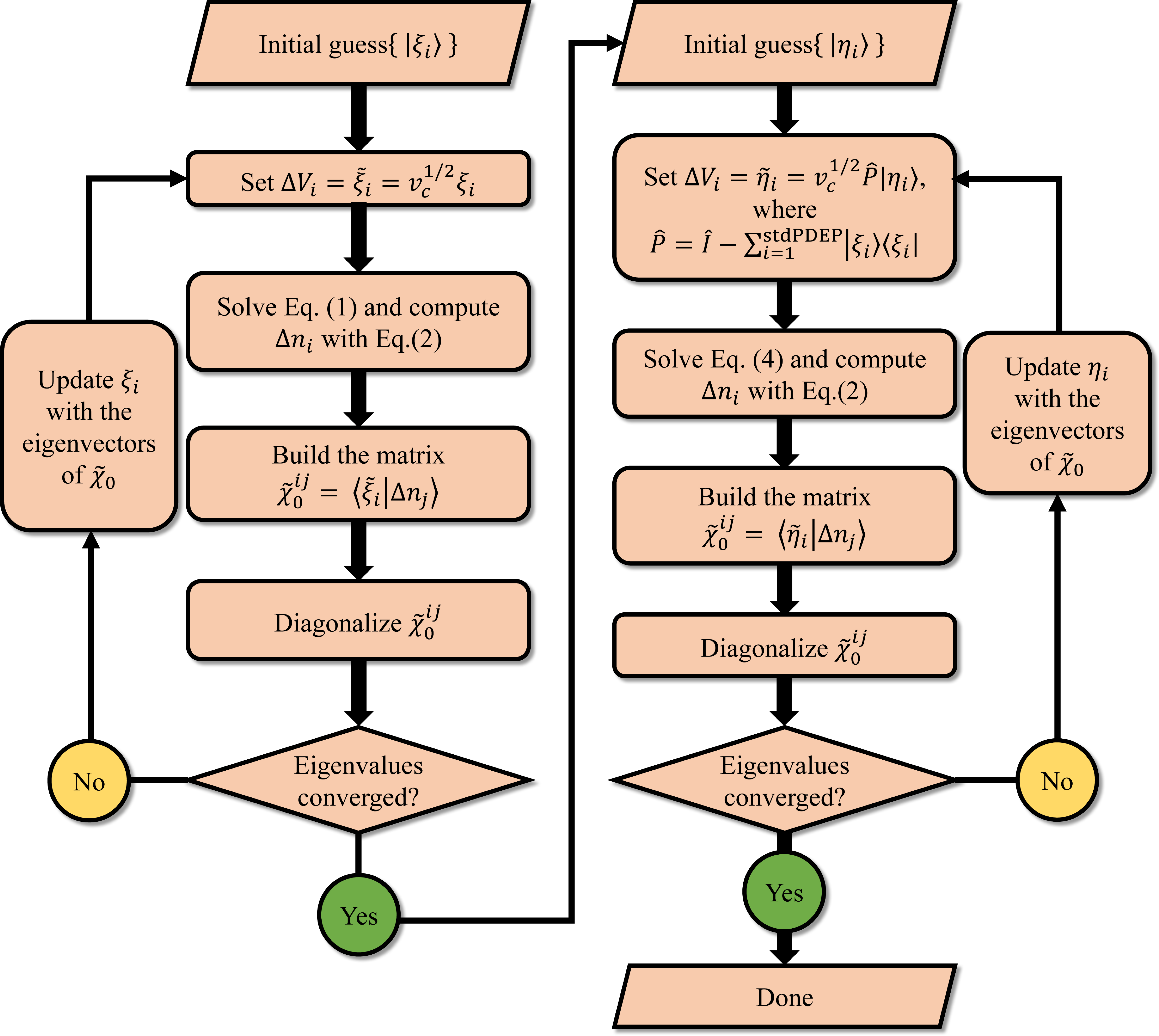}
    \caption{The workflow used in this work to generate eigenvectors of the dielectric matrix using the Kohn-Sham Hamiltonian (stdPDEP) and using the kinetic operator (kinPDEP). See text.}
    \label{fig:workflow}
\end{figure}

In the following, we refer to the eigenpotentials from \autoref{equ:Sternheimer} as standard PDEPs (stdPDEP, $\xi_i,\,\,i = 1, \cdots, N_\mathrm{stdPDEP}$) and those from \autoref{equ:KineticSternheimer} as kinetic PDEPs (kinPDEP, $\eta_i,\,\,i = 1, \cdots, N_\mathrm{kinPDEP}$) and we rewrite the irreducible density-density response function as

\begin{equation}
    \tilde{\chi}_0 = \sum_{i=1}^{N_\mathrm{stdPDEP}}\left|\xi_i\right\rangle\lambda_i\left\langle\xi_i\right|+\sum_{j=1}^{N_\mathrm{kinPDEP}}\left|\eta_j\right\rangle\mu_j\left\langle\eta_j\right|,
\end{equation}
where $\xi_i$ and $\eta_j$ are standard and kinetic PDEPs, respectively, and $\lambda_i$ and $\mu_j$ are their corresponding eigenvalues. The procedure to generate stdPDEPs and kinPDEPs is summarized in \autoref{fig:workflow}. We note that during the construction of kinetic PDEPs, the projection operator $\hat{P}=\hat{I}-\sum_{N_\mathrm{stdPDEP}}\left|\xi_i\right\rangle\left\langle\xi_i\right|$ was applied so as to satisfy the orthonormality constraint, $\langle\xi_i|\xi_j\rangle = \delta_{ij}$, $\langle\eta_i|\eta_j\rangle=\delta_{ij}$ and $\langle\xi_i|\eta_j\rangle=0$, $\forall (i,j)$; in addition we applied $v_c^{1/2}$ to perturbations to yield a \textit{symmetrized} irreducible response function $\tilde{\chi}_0$.

In our $G_0W_0$ calculations, both the static Green's function and the statically screened Coulomb interaction are written in the basis of eigenpotentials of the dielectric matrix. Frequency integration is performed using a contour deformation algorithm. A detailed description of  the implementation of $G_0W_0$ calculations in the basis of eigenpotentials can be found in Ref. \citenum{Nguyen2012-GW,Anh2013-GW,Marco2015}. One should also note that \autoref{equ:Sternheimer} and \ref{eq:deltadensity} apply only to semiconductors, but this formalism may be generalized to metallic systems.

\section{\label{sec:results}Validation and results}
We now turn to discussing results for molecules and solids obtained by using a combination of standard and kinetic PDEPs. To examine the efficiency and applicability of the approximation proposed in \autoref{sec:method}, we performed $G_0W_0$ calculations for a set of closed-shell small molecules, a larger molecule (Buckminsterfullerene $\mathrm{C_{60}}$), and an amorphous silicon nitride/silicon interface ($\mathrm{Si_3N_4/Si}$) with a total of 1152 valence electrons. All Kohn-Sham eigenvalues and eigenvectors were obtained with the QuantumEspresso package,\cite{QE2009,QE2017} using the PBE approximation\cite{PBEfunctional}, SG15\cite{ONCV2015} ONCV\cite{hamann2013ONCV} pseudopotentials and $G_0W_0$ calculations were carried out with the WEST code.\cite{Marco2015}

\begin{figure}
    \centering
    \includegraphics[width=0.6\linewidth]{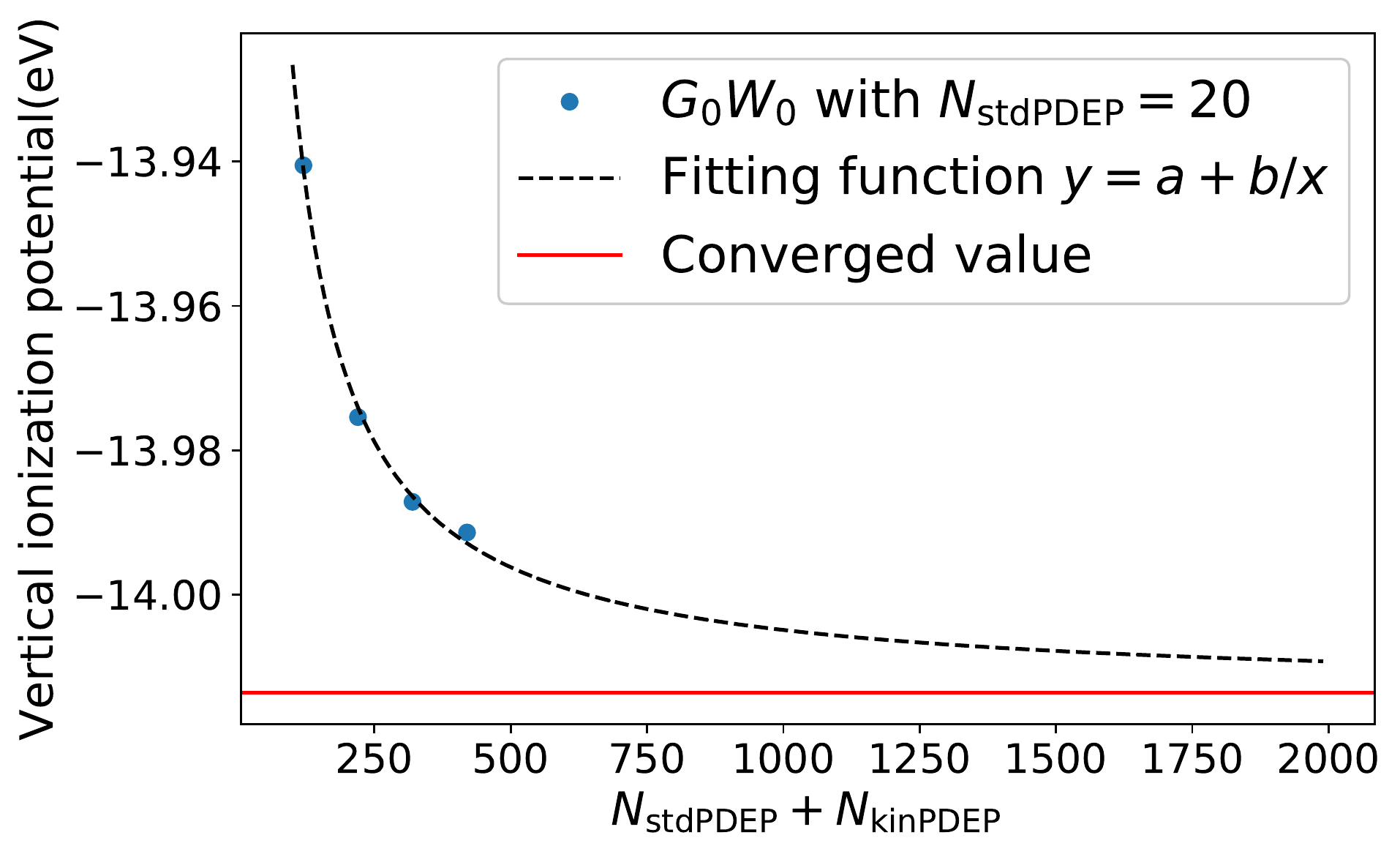}
    \caption{Extrapolation of $G_0W_0$ energy of highest occupied orbital of methane with respect to total number of eigenpotential used ($N_\mathrm{stdPDEP}+N_\mathrm{kinPDEP}$). In this plot, $N_\mathrm{stdPDEP} = 20$ and $N_\mathrm{kinPDEP} = 100,\,200,\,300,\,400$ for the four points, respectively.}
    \label{fig:fitting_example}
\end{figure}

We first considered a subset of molecules taken from the G2/97 test set\cite{G2/97--testset} and calculated their vertical ionization potential (VIP) and electron affinity (EA) using different numbers of stdPDEPs ($N_\mathrm{stdPDEP}$) and kinPDEPs ($N_\mathrm{kinPDEP}$). We chose a plane wave cutoff of 85 Ry and a periodic box of edge 30 Bohr. For all molecules, we included either 20 or 100 stdPDEPs in our calculations, then added 100, 200, 300, 400 kinPDEPs in subsequent calculations, after which an extrapolation was performed ($a+\frac{b}{N_\mathrm{stdPDEP}+N_\mathrm{kinPDEP}}$) to find converged values (one example is shown in \autoref{fig:fitting_example}). These results are given in the second (A) and third columns (B) of \autoref{tab:VIP-PBE-small} and \autoref{tab:EA-PBE-small}.\footnote{Here we did not attempt to solve the problem on how to accurately compute resonant molecular energy levels: our goal is to compare results obtained with solutions of the Sternheimer equation using the full Hamiltonian (\autoref{equ:Sternheimer}) and approximate solutions using only the kinetic operator (\autoref{equ:KineticSternheimer}). As long as the results obtained with the two procedures agree, we consider the results obtained from \autoref{equ:KineticSternheimer} as accurate.} The reference results reported in the last column (C) of the two tables were obtained with 200, 300, 400, 500 stdPDEPs and an extrapolation was applied. We found that including only 20 stdPDEPs yields quasiparticle energies accurate within $0.1\,\mathrm{eV}$ relative to the reference $G_0W_0$ values obtained using only standard eigenpotentials. (See also Fig. 2 of the Supplementary Information). The two data sets starting from 20 or 100 stdPDEPs enabled us to save 40\% and 10\% of computer time compared to the time usage needed with only standard eigenpotentials.

\begin{table}
    \centering
    \caption{Vertical ionization potential (eV) obtained at the $G_0W_0@\mathrm{PBE}$ level of theory with different numbers of standard and kinetic PDEPs. (A) 20 stdPDEPs + up to 400 kinPDEPs and extrapolated; (B) 100 stdPDEPs  + up to 400 kinPDEPs and extrapolated; (C) pure stdPDEPs and extrapolated. A detailed discussion of extrapolations of quasiparticle energies can be found in Ref. \citenum{Marco2018}. See also  Fig. 3 of the SI.\label{tab:VIP-PBE-small}}
    \begin{tabular}{cccc}
    \hline\hline
    Molecule & A & B & C \\
    \hline
  $\mathrm{C_2H_2}$ &  11.07 &  11.06 &  11.06 \\
  $\mathrm{C_2H_4}$ &  10.41 &  10.40 &  10.40 \\
 $\mathrm{C_4H_4S}$ &   8.80 &   8.77 &   8.76 \\
  $\mathrm{C_6H_6}$ &   9.17 &   9.14 &   9.13 \\
 $\mathrm{CH_3Cl}$  &  11.28 &  11.26 &  11.25 \\
 $\mathrm{CH_3OH}$  &  10.58 &  10.56 &  10.56 \\
 $\mathrm{CH_3SH}$  &   9.39 &   9.36 &   9.36 \\
   $\mathrm{CH_4}$  &  14.01 &  14.01 &  14.01 \\
   $\mathrm{Cl_2}$  &  11.51 &  11.51 &  11.50 \\
   $\mathrm{ClF}$   &  12.55 &  12.55 &  12.54 \\
    $\mathrm{CO}$   &  13.51 &  13.50 &  13.50 \\
   $\mathrm{CO_2}$  &  13.32 &  13.31 &  13.31 \\
    $\mathrm{CS}$   &  11.00 &  10.98 &  10.98 \\
    $\mathrm{F_2}$  &  14.99 &  14.97 &  14.97 \\
  $\mathrm{H_2CO}$  &  10.43 &  10.42 &  10.42 \\
   $\mathrm{H_2O}$  &  11.82 &  11.82 &  11.81 \\
  $\mathrm{H_2O_2}$ &  10.87 &  10.87 &  10.86 \\
   $\mathrm{HCl}$   &  12.50 &  12.50 &  12.50 \\
   $\mathrm{HCN}$   &  13.20 &  13.20 &  13.20 \\
   $\mathrm{Na_2}$  &   4.95 &   4.95 &   4.95 \\
   \hline\hline
    \end{tabular}
\end{table}

\begin{table}
    \centering
    \caption{Vertical electron affinity (eV) obtained at the $G_0W_0@\mathrm{PBE}$ level of theory with different numbers of standard and kinetic PDEPs. (A) 20 stdPDEPs + up to 400 kinPDEPs and extrapolated; (B) 100 stdPDEPs  + up to 400 kinPDEPs and extrapolated; (C) pure stdPDEPs and extrapolated. We defined the electron affinity as the energy of the first unoccupied state, within the given unit cell used, extrapolated as a function of $N_\mathrm{stdPDEP} + N_\mathrm{kinPDEP}$. See also Ref. \citenum{Marco2018} and Fig. 3 of the SI.\label{tab:EA-PBE-small}}
    \begin{tabular}{cccc}
    \hline\hline
    Molecule & A & B & C \\
    \hline
  $\mathrm{C_2H_2}$ &  -2.42 &  -2.41 &  -2.41 \\
  $\mathrm{C_2H_4}$ &  -1.75 &  -1.75 &  -1.75 \\
 $\mathrm{C_4H_4S}$ &  -0.85 &  -0.81 &  -0.80 \\
  $\mathrm{C_6H_6}$ &  -1.01 &  -0.96 &  -0.96 \\
 $\mathrm{CH_3Cl}$  &  -1.17 &  -1.16 &  -1.16 \\
 $\mathrm{CH_3OH}$  &  -0.89 &  -0.89 &  -0.89 \\
 $\mathrm{CH_3SH}$  &  -0.88 &  -0.88 &  -0.88 \\
   $\mathrm{CH_4}$  &  -0.64 &  -0.64 &  -0.64 \\
   $\mathrm{Cl_2}$  &   1.65 &   1.64 &   1.65 \\
   $\mathrm{ClF}$   &   1.28 &   1.28 &   1.28 \\
    $\mathrm{CO}$   &  -1.56 &  -1.57 &  -1.57 \\
   $\mathrm{CO_2}$  &  -0.97 &  -0.97 &  -0.97 \\
    $\mathrm{CS}$   &   0.49 &   0.51 &   0.51 \\
    $\mathrm{F_2}$  &   1.16 &   1.16 &   1.16 \\
  $\mathrm{H_2CO}$  &  -0.69 &  -0.68 &  -0.68 \\
   $\mathrm{H_2O}$  &  -0.90 &  -0.90 &  -0.90 \\
  $\mathrm{H_2O_2}$ &  -1.80 &  -1.79 &  -1.79 \\
   $\mathrm{HCl}$   &  -1.07 &  -1.07 &  -1.07 \\
   $\mathrm{HCN}$   &  -2.08 &  -2.08 &  -2.08 \\
   $\mathrm{Na_2}$  &   0.64 &   0.63 &   0.63 \\
   \hline\hline
    \end{tabular}
\end{table}

	\autoref{tab:C60} shows our results for the $\mathrm{C_{60}}$ molecule. The structure of $\mathrm{C_{60}}$ (point group $I_h$) was also taken from the NIST computational chemistry database,\cite{NIST-CCCBDB} (optimized with the $\omega$B97X-D functional and cc-pVTZ basis sets) and no further optimization was carried out. We used the PBE exchange-correlation functional, a plane wave energy cutoff of $\SI{40}{\si{Ry}}$ and cell size of $\SI{40}{\si{bohr}}$, the same as used in Ref. \citenum{Anh2013-GW}. We performed two groups of calculations starting with 100 and 200 standard eigenpotentials, respectively. For both calculations, we computed quasiparticle energies by adding 100, 200, 300, 400 kinetic eigenpotentials and extrapolation was done in the same manner. As seen in \autoref{tab:C60}, the results obtained with 200 standard eigenpotentials and additional kinetic eigenpotentials differ at most by \SI{0.1}{\si{eV}} from those computed with standard eigenpotentials (extrapolated up to $N_\mathrm{stdPDEP}=2000$). The two sets of calculations starting with $N_\mathrm{stdPDEP}=100$ and $N_\mathrm{kinPDEP}=200$ amounted to savings of 32\% and 15\%, respectively.

\begin{figure}
		\centering
		\includegraphics[width=0.6\textwidth]{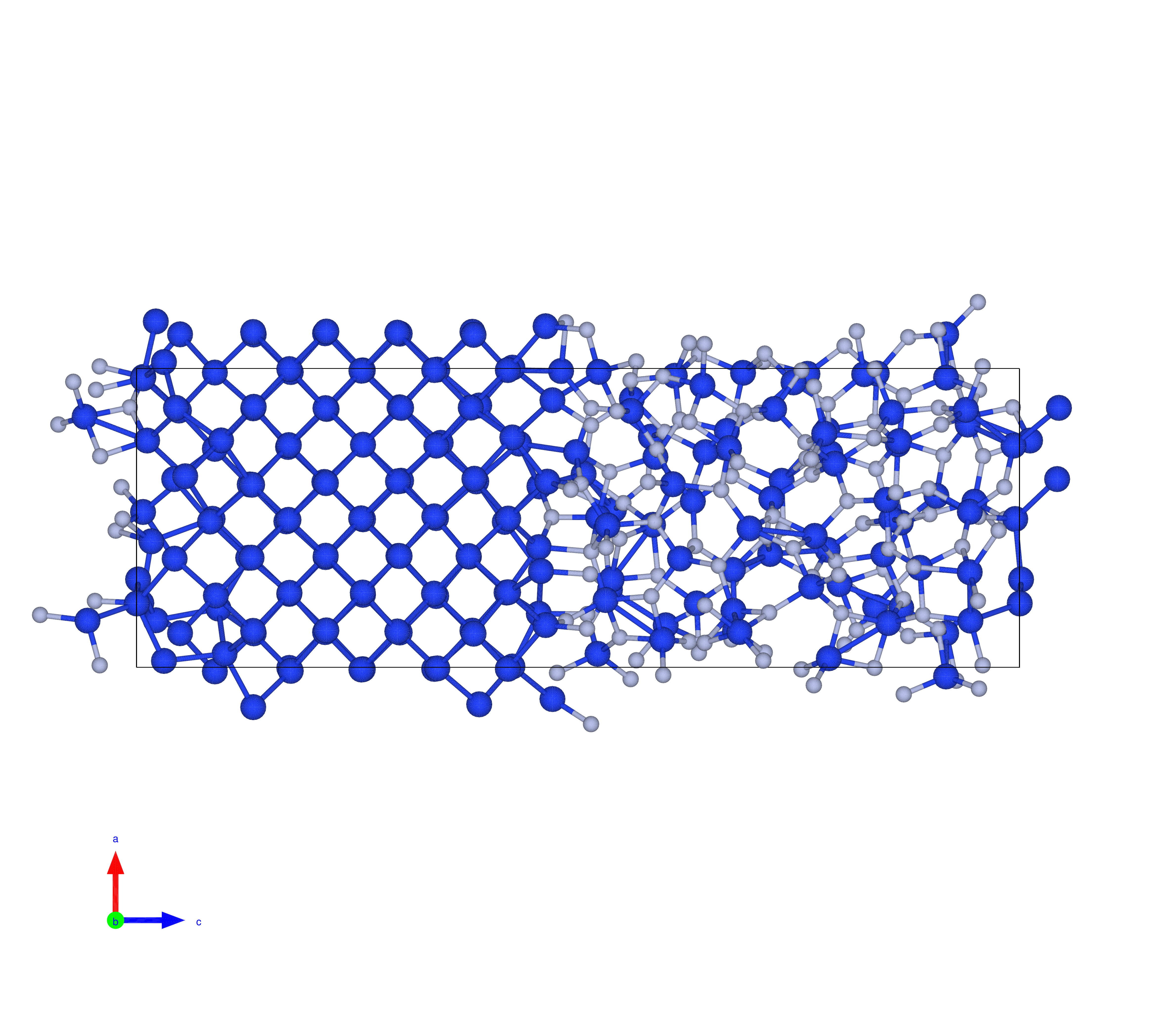}
		\caption{Ball and stick representation of the atomistic structure\cite{Anh2013interface} of the $\mathrm{Si_3N_4/Si(100)}$ interface used in our study.}\label{fig:interface_structure}
	\end{figure}

	\begin{table}
		\noindent\makebox[\textwidth]{
		\begin{threeparttable}
			\centering
			\caption{Quasiparticle energies (eV) of $\mathrm{C_{60}}$ calculated at the $G_0W_0$@PBE level of theory. Energy levels are labeled by their symmetry in point group $I_h$. $N_\mathrm{stdPDEP}=100$ and $N_\mathrm{stdPDEP}=200$ are calculations with 100 and 200 stdPDEPs and up to 400 kinPDEPs and extrapolated. $N_\mathrm{kinPDEP} = 0$ is the calculation with pure stdPDEPs and extrapolated. (See text)}\label{tab:C60}
			\small
			\begin{tabular}{cccccc}
			\hline\hline
				Energy levels& $N_\mathrm{stdPDEP}=100$ & $N_\mathrm{stdPDEP}=200$ & $N_\mathrm{kinPDEP} = 0$ & $G_0W_0$ & Expt \\
			\hline
			$t_{1g}$ & -1.70 & -1.66 & -1.70 & & \\
			$t_{1u}$ & -2.70 & -2.77 & -2.82 & -2.74\tnote{a}, -2.62\tnote{b}, -2.82\tnote{c} & -2.69\tnote{d} \\
			$h_u$ & -7.32 & -7.32 & -7.38 & -7.31\tnote{a}, -7.21\tnote{b}, -7.37\tnote{c} & -7.61\tnote{d}, -7.6\tnote{e}\\
			$g_g+h_g$ & -8.46, -8.52 & -8.46, -8.51 & -8.51, -8.56 & -8.68\tnote{c}, -8.69\tnote{c} & -8.59\tnote{e} \\
			\hline\hline
			\end{tabular}
			\begin{tablenotes}
				\item[a] Ref. \citenum{Anh2013-GW}: with 700 standard eigenpotentials;
				\item[b] Ref. \citenum{Louie2011PRL}: with 27387 SAPOs, where SAPO stands for simple approximate physical orbitals;
				\item[c] Ref. \citenum{Marzari2015C60}: plane wave energy cutoff of $\SI{45}{\si{Ry}}$ and cell edge of $\SI{31.7}{\si{bohr}}$;
				\item[d] Ref. \citenum{NIST-CHEM-BOOK};
				\item[e] Ref. \citenum{C60-expt-1991}.
			\end{tablenotes}
		\end{threeparttable}
		}
	\end{table}
	We now turn to a more complex system, amorphous silicon nitride interfaced with a silicon surface ($\mathrm{Si_3N_4/Si(100)}$), whose structure was taken from Ref. \citenum{Anh2013interface} (See \autoref{fig:interface_structure}). This interface is representative of a heterogeneous, low dimensional system.

	We computed band offsets (BO) by employing two different methods. The first one is based on the calculation of the local density of electronic states (LDOS);\cite{Anh2013interface,LDOS-cited-by-Anh} the second one is based on the calculation of the average electrostatic potential which is then used to set a common zero of energy on the two parts of the slab representing the two solids interfaced with each other.\cite{Walle-alignment-by-potential} The average electrostatic potential was fitted with the method proposed in Ref. \citenum{YuanPing2017potential}. We used a plane wave energy cutoff of $\SI{70}{\si{Ry}}$. We also performed $G_0W_0$@PBE calculations for each bulk system separately and obtained quasiparticle energies.

	The local density of states is given by:
	\begin{equation}
		D(\varepsilon,z) = 2\sum_i\int\frac{\mathrm{d}x}{L_x}\int\frac{\mathrm{d}y}{L_y} |\psi_i(x,y,z)|^2\delta(\varepsilon-\varepsilon_i),
	\end{equation}
	where $z$ is the direction perpendicular to the interface, $\psi_i(x,y,z)$ is the wavefunction, the factor 2 represents spin degeneracy. We computed the variation of the valence band maximum(VBM) and conduction band minimum(CBM) as a function of the direction ($z$) perpendicular to the interface\cite{Anh2013interface}
	\begin{equation}
		\int_\mathrm{VBM}^{E_F} D(\varepsilon,z)\,\mathrm{d}\varepsilon = \int_{E_F}^\mathrm{CBM}D(\varepsilon,z)\,\mathrm{d}\varepsilon = \Delta \int_{-\infty}^{E_F}D(\varepsilon,z)\,\mathrm{d}\varepsilon
	\end{equation}
	where $E_F$ is the Fermi energy and $\Delta$ is an constant that is chosen to be $0.003$.\cite{Anh2013interface} We follow a common procedure adopted to describe the electronic structure of interfaces described in Ref. \citenum{Anh2013interface} and \citenum{LDOS-cited-by-Anh}.
 The band offsets (see \autoref{tab:bandoffsets}) at the PBE level of theory were determined to be $\SI{0.83}{\si{eV}}$ and $\SI{1.49}{\si{eV}}$ for the valence band and conduction band, respectively, which are in agreement with the results of $\SI{0.8}{\si{eV}}$ and $\SI{1.5}{\si{eV}}$ reported in Ref. \citenum{Anh2013interface}.

	As mentioned above, another method to obtain the valence band offset (VBO) and conduction band offset (CBO) is to align energy levels with respect to electrostatic potentials. Following Ref. \citenum{YuanPing2017potential}, the electrostatic potential was computed as:
	\begin{equation}
		\bar{V}(\mathbf{r}) = V_H(\mathbf{r}) + V_\mathrm{loc}(\mathbf{r}) - \sum_i\bar{V}_\mathrm{at}^{(i)}(|\mathbf{r}-\mathbf{r}_i|)
	\end{equation}
	where $\bar{V}_\mathrm{at}^{(i)}$ is the potential near the core region obtained from neutral atom calculations. With this method, VBO and CBO at the PBE level are found to be $\SI{0.89}{\si{eV}}$ and $\SI{1.63}{\si{eV}}$.

	To compute $G_0W_0$ corrections on band offsets, we performed $G_0W_0$@PBE calculations of bulk silicon and amorphous silicon nitride. In \autoref{tab:Bulk-QP}, quasiparticle corrections to Kohn-Sham energies of bulk silicon and amorphous silicon nitride are shown. The second and third columns are computed with 1000 and 2000 standard eigenpotentials. The fitted $G_0W_0$ reference results are extrapolated with 500, 1000, 1500 and 2000 standard eigenpotentials. To test accuracy of kinetic eigenpotentials, we started with 400 stdPDEPs and added 100, 200, 300, 400 kinPDEPs, after which the same extrapolation was applied. We calculated VBO and CBO at the $G_0W_0$ level by applying quasiparticle corrections on PBE results. After applying quasiparticle corrections on LDOS results, VBO and CBO are $\SI{1.41}{\si{eV}}$ and $\SI{1.88}{\si{eV}}$ while the VBO and CBO are found to be $\SI{1.46}{\si{eV}}$ and $\SI{2.02}{\si{eV}}$ after applying corrections to results based on electrostatic potential alignment. Both of them are close to the range of experimental results of $1.5-\SI{1.78}{\si{eV}}$ and $1.82-\SI{2.83}{\si{eV}}$. Time saving when using kinetic eigenpotentials to obtain quasiparticle corrections was approximately $\sim 50\%$.

    \begin{table}
        \centering
        \caption{Quasiparticle energies of valence band maximum (VBM) and conduction band minimum (CBM) of bulk silicon and amorphous $\mathrm{Si_3N_4}$ computed with standard eigenpotentials and by combining standard and kinetic eigenpotentials. Columns $N_\mathrm{stdPDEP}=1000$ and $N_\mathrm{stdPDEP}=2000$ are calculations done with 1000 and 2000 stdPDEPs; column Fit is extrapolated results; column $N_\mathrm{kinPDEP} = 400$ is calculation with up to 400 kinPDEPs and extrapolated. (See text) }
        \begin{tabular}{c|ccccc}
        \hline\hline
                          & $N_\mathrm{stdPDEP}=1000$ & $N_\mathrm{stdPDEP}=2000$ & Fit & $N_\mathrm{kinPDEP} = 400$ & \\
        \hline
            $\mathrm{Si}$ VBM        & 5.70  & 5.55  & 5.45  &  5.53 \\
            $\mathrm{Si}$ CBM        & 7.03  & 6.91  & 6.79  &  6.82 \\
            $a-\mathrm{Si_3N_4}$ VBM & 7.14  & 7.01  & 7.01  &  6.99 \\
            $a-\mathrm{Si_3N_4}$ CBM & 11.99 & 11.87 & 11.83 & 11.83 \\
        \hline\hline
        \end{tabular}
        \label{tab:Bulk-QP}
    \end{table}

	\begin{table}
		\centering
		\begin{threeparttable}
		\caption{Band gaps of bulk Si, $a-\mathrm{Si_3N_4}$, and band offsets (VBO\&CBO) of the interface.(see \autoref{fig:interface_structure}) All values are in $\si{eV}$.}
		\label{tab:bandoffsets}
		\begin{tabular}{cc|cccc}
			\hline\hline
			 \multicolumn{2}{c|}{\diagbox{Method}{Energy}} 	& VBO & CBO & $E_g^\mathrm{Si}$ & $E_g^\mathrm{Si_3N_4}$ \\
			\hline
			\multirow{3}{*}{PBE} 				& LDOS      		& 0.83 & 1.49 & 0.67 & 3.19 \\
													 				& Potential 		& 0.89 & 1.63 & 0.76 & 3.19 \\
													 				& Ref\tnote{a}  & 0.8  & 1.5  & 0.7  & 3.17 \\
			\hline
			\multirow{3}{*}{$G_0W_0$}		& LDOS  				& 1.41 & 1.88 & 1.29 & 4.77 \\
																	& Potential 		& 1.46 & 2.02 & 1.29 & 4.77 \\
																	& Ref\tnote{a} 	& 1.5  & 1.9  & 1.3  & 4.87 \\
			\hline
			\multicolumn{2}{c|}{Expt}               		& 1.5-1.78\tnote{b} & 1.82-2.83\tnote{c} & 1.17\tnote{d} & 4.5-5.5\tnote{e} \\
			\hline\hline
		\end{tabular}
		\begin{tablenotes}
			\item[a] Ref. \citenum{Anh2013interface};
			\item[b] Ref. \citenum{Expt1,Expt2,Expt3};
			\item[c] Estimated by the other three experimental values;
			\item[d] Ref. \citenum{Expt4};
			\item[e] Ref. \citenum{Expt5,Expt6,Expt7}.
		\end{tablenotes}
	\end{threeparttable}
	\end{table}

\section{Conclusion\label{sec:conclusion}}
The method introduced in Ref. \citenum{Nguyen2012-GW,Anh2013-GW,Marco2015,Marco2018} to compute quasiparticle energies using the $G_0W_0$ approximation avoids the calculation of virtual electronic states and the inversion and storage of large dielectric matrices, thus leading to substantial computational savings. Building on the strategy proposed in Ref. \cite{wilson2008,wilson2009} and implemented in the WEST code,\cite{Marco2015} here we proposed an approximation of the spectral decomposition of dielectric matrices that further improve the efficiency of $G_0W_0$ calculations. In particular we built sets of eigenpotentials used as a basis to expand the Green function and the screened Coulomb interaction by solving two separate Sternheimer equations: one using the Hamiltonian of the system, to obtain the eigenvectors corresponding to the lowest eigenvalues of the response function,  and one equation using just the kinetic energy operator to obtain the eigenpotentials corresponding to higher eigenvalues.  We showed that without compromising much accuracy, this approximation reduces the cost of $G_0W_0$ calculations by 10\%-50\%, depending on the system, with the most savings observed for the largest systems studied here.

\section*{Supplementary Material}
See supplementary material for convergence studies of the $G_0W_0$ calculations of the vertical ionization potential of the $\mathrm{CH_4}$ molecule, whcih is taken as a representative example of the molecular systems studied in the main text.

\begin{acknowledgments}
We thank He Ma, Ryan L. McAvoy and Ngoc Linh Nguyen for useful discussions. This work was supported by MICCoM, as part of the
Computational Materials Sciences Program funded by the U.S.
Department of Energy, Office of Science, Basic Energy
Sciences, Materials Sciences and Engineering Division through
Argonne National Laboratory, under contract number DE-AC02-06CH11357. This research used resources of the Argonne Leadership Computing Facility, which is a DOE Office of Science User Facility supported under Contract DE-AC02-06CH11357, and resources of the University of Chicago Research Computing Center.
\end{acknowledgments}

\clearpage
\bibliography{main}

\end{document}


\title{Supplementary Information:
Improving the efficiency of $G_0W_0$ calculations with approximate spectral decompositions of dielectric matrices: Supplementary information}
\renewcommand{\figureautorefname}{Fig.}

\author{Han Yang}
\affiliation{Department of Chemistry, University of Chicago, Chicago, Illinois 60637, United States}
\affiliation{Pritzker School of Molecular Engineering, University of Chicago, Chicago, Illinois 60637, United States}

\author{Marco Govoni}
\affiliation{Pritzker School of Molecular Engineering, University of Chicago, Chicago, Illinois 60637, United States}
\affiliation{Materials Science Division and Center for Molecular Engineering, Argonne National Laboratory, Lemont, Illinois 60439, United States}

\author{Giulia Galli}
\email[]{gagalli@uchicago.edu}
\affiliation{Department of Chemistry, University of Chicago, Chicago, Illinois 60637, United States}
\affiliation{Pritzker School of Molecular Engineering, University of Chicago, Chicago, Illinois 60637, United States}
\affiliation{Materials Science Division and Center for Molecular Engineering, Argonne National Laboratory, Lemont, Illinois 60439, United States}

\date{\today}

\maketitle

This Supplementary Information reports convergence studies of the G$_0$W$_0$ calculations of the vertical ionization potential of the CH$_4$ molecule, whcih is taken as a representative example of the molecular systems studied in the main text.

\section*{$G_0W_0$ calculations of the vertical ionization potential of the methane molecule}
\subsection{Eigenvalues of the symmetrized \textit{irreducible} density-density response function}
\begin{figure}
    \centering
    \includegraphics[width=0.8\linewidth]{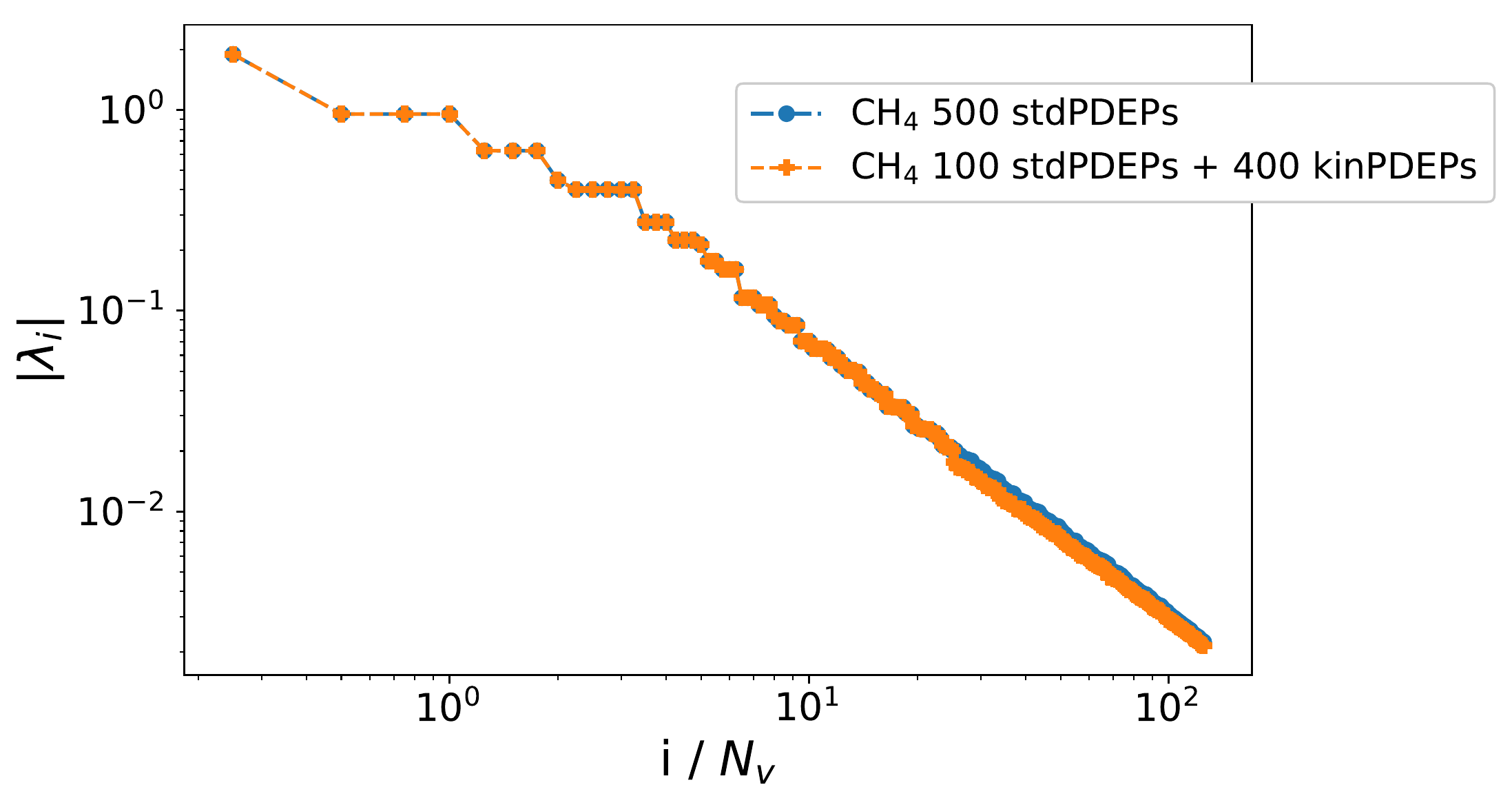}
    \caption{Comparison between the eigenvalues ($\lambda_i$) of the leading 500 stdPDEPs and the eigenvalues of the 100 leading stdPDEPs followed by 400 kinPDEPs of the CH4 molecule. $N_v$ is the number of occupied states and stdPDEPs and kinPDEPs are eigenvectors of the symmetrized irreducible density-density response function($\tilde{\chi}_0$) solved using Kohn-Sham Hamiltonian and kinetic operator.}
    \label{fig:compare_eigenvals}
\end{figure}

In \autoref{fig:compare_eigenvals}, we  compare eigenvalues for the symmetrized irreducible density-density response function of the methane molecule: 500 stdPDEPs and 100 stdPDEPs +  400 kinPDEPs. On the scale of the figure the results are indistinguishable.

\subsection{Calculations of the vertical ionization potential}
\begin{figure}
    \centering
    \includegraphics[width=0.8\linewidth]{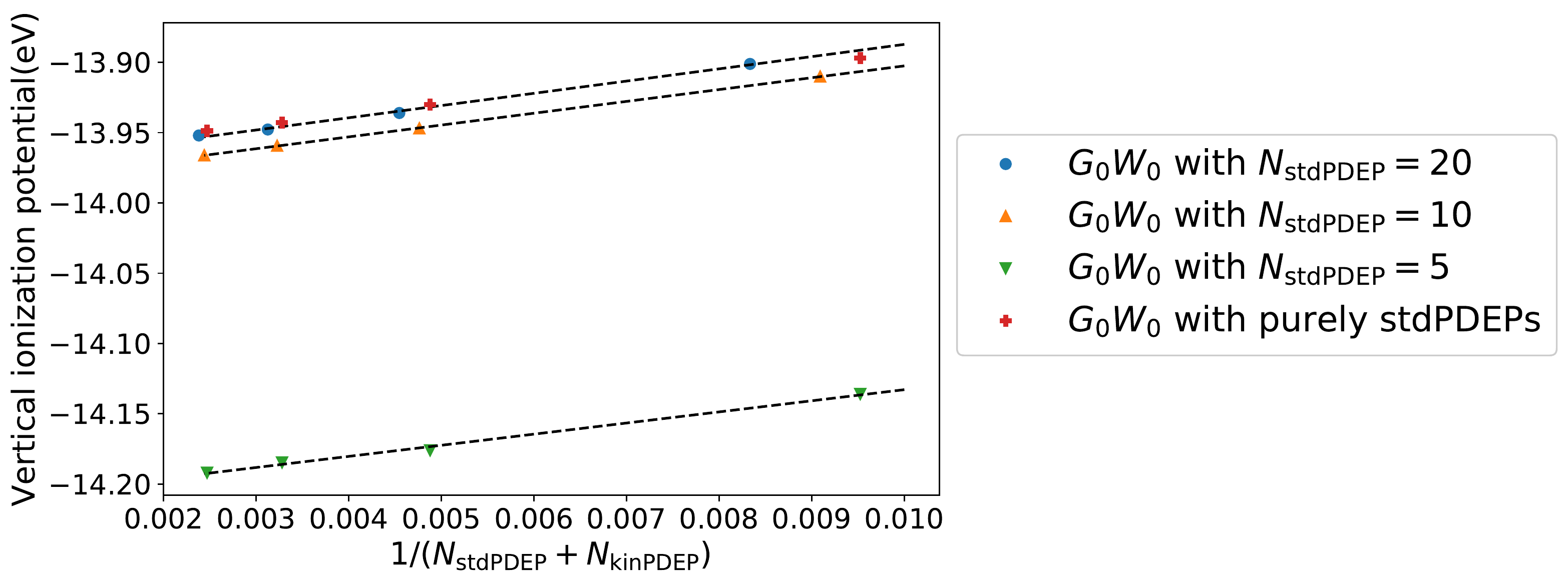}
    \caption{Calculations of the vertical ionization potential of the methane molecule with 20, 10, 5 standard eigenpotentials (stdPDEP) and up to 400 kinetic eigenpotentials (kinPDEP) compared to calculations (red symbols) performed  with purely stdPDEPs.}
    \label{fig:different_number_of_PDEPs}
\end{figure}

We present, in \autoref{fig:different_number_of_PDEPs}, the results for the vertical ionization potential of the $\mathrm{CH_4}$ molecule computed with 5, 10, 20 standard PDEPs ($N_\mathrm{stdPDEP}$), and the remaining 100, 200, 300 and 400 PDEPs treated as kinetic PDEPs. When setting $N_\mathrm{stdPDEP}$ = 10 or 20 we obtain results accurate within 0.02 eV, as compared to the ones obtained using only standard PDEP. When using 5 stdPDEPs we obtain instead an error more than 10 times larger (0.25 eV). In terms of computational savings, we save 35\% with 20 stdPDEPs and 40\% with 5 stdPDEPs for the methane molecule.

\subsection{Interpolation of results}
In this work, the computed energy levels  were  interpolated with respect to the total number of PDEPs,
\begin{equation}
    E = a + \frac{b}{N_\mathrm{stdPDEP}+N_\mathrm{kinPDEP}},
\end{equation}
where $a$ is the converged energy level and $b$ is an arbitrary number depending on the system.

\begin{figure}
    \centering
    \includegraphics[width=0.8\linewidth]{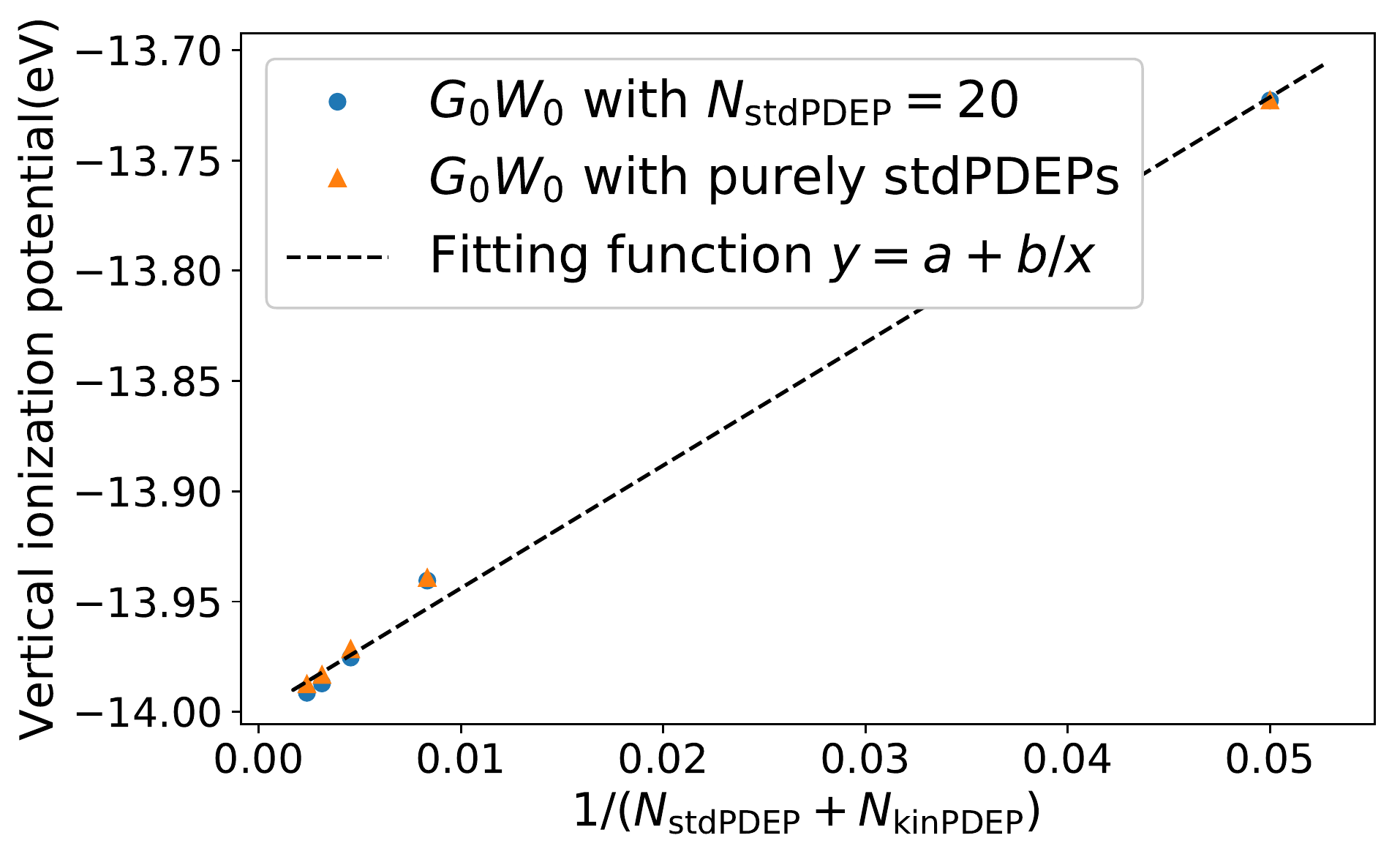}
    \caption{Interpolation of the computed vertical ionization potential of the methane molecule with respect to the total number of PDEPs, where PDEPs are eigenvectors of the symmetrized irreducible density-density response function ($\tilde{\chi}_0$).}
    \label{fig:interpolation_from20}
\end{figure}

In \autoref{fig:interpolation_from20}, we show results for the vertical ionization potential  of the methane molecule with respect to the total number of PDEPs included in the calculation: 20 stdPDEPs were used followed by 0, 100, 200, 300, 400 kinPDEPs. We note that as the number of  kinPDEPs increases, one obtains energies similar to those computed with stdPDEPs; in addition the proportionality between energies and the inverse of the total number of PDEPs holds even when the total number of PDEPs in small.

